\documentclass[aps,showpacs,bibnotes]{revtex4}
\usepackage{graphicx}
\begin{document}
\draft
\def\ds{\displaystyle}
\title{Off-axis Vortex in a Rotating Dipolar Bose-Einstein Condensation}
\author{C. Yuce, Z. Oztas}
\address{ Physics Department, Anadolu University,
 Eskisehir, Turkey}
\email{cyuce@anadolu.edu.tr}
\date{\today}
\begin{abstract}
We consider a singly quantized off-axis straight vortex in a
rotating dipolar ultracold gas in the Thomas-Fermi (TF) regime. We
derive analytic results for small displacements and perform
numerical calculations for large displacement within the TF
regime. We show that the dipolar interaction energy increases
(decreases) as the vortex moves from the trap center to the edge
in an oblate (a prolate) trap. We find that for an oblate (a
prolate) trap, the effect of the dipole-dipole interaction is to
lower (raise) both the precession velocity of an off-center
straight vortex line and the angular velocity representing the
onset of metastability.
\end{abstract}
\maketitle

\section{Introduction}

The first experimental detection of a vortex in a dilute
alkali-atomic gas Bose-Einstein condensate (BEC) was made by
Matthews et al. in 1999 using $\ds{^{87}Rb}$ atoms
\cite{ilkdeney1} and theoretical predictions on the main features
of the vortex states have been shown to agree with experiments
\cite{theory} (references therein). $\ds{^{87}Rb}$ has a small
dipole moment while chromium atoms posses a larger permanent
magnetic dipole moment, which leads to significant dipolar
interactions in addition to the usual short-range interactions.
The successful Bose-Einstein condensation of $\ds{^{52}Cr}$ atoms
has stimulated a growing interest in the study of BEC with
nonlocal dipole-dipole interactions
\cite{deney01,deney02,deney03}. This nonlocal character has
remarkable consequences for the physics of rotating dipolar gases
\cite{TFvortex,genel0,genel1,genel2,genel3,genel5,genel6,genel7,genel8,genel4,genel9,genel10,vortex2,vortex1,vortexek}.
It has been shown that the critical angular frequency for vortex
creation may be significantly affected by the dipolar interaction
\cite{TFvortex}. In addition, dipolar gases under fast rotation
develop vortex lattices, which due to the dipolar interaction may
be severely distorted \cite{genel6}, and even may change its
configuration from the usual triangular Abrikosov lattice into
other arrangements \cite{genel7,genel8}. It was shown that, the
dipolar interaction may significantly modify the vortex line
stability. Under appropriate conditions, the dispersion law for
transverse modes shows a rotonlike minimum, which for sufficiently
large dipolar interaction may reach zero energy, destabilizing the
Kelvin waves \cite{genel4}. In the TF limit, dipole-dipole
interaction changes the stability and the instability conditions
and the possibility of vortex lattice formation for a rotating
dipolar BEC in an elliptical trap \cite{vortex2}.\\
The long-range and anisotropic interactions introduce rich
physical effects, as well as opportunities to control BECs. In a
prolate dipolar gas with the dipoles polarized along the z-axis,
the dipolar interaction is attractive, whereas it is repulsive for
an oblate dipolar gas. As a result, the sign of the dipolar
mean-field energy can be controlled via the trap aspect ratio. In
this paper, we will consider an off-axis vortex lines in an oblate
dipolar BEC with the dipoles aligned in the z direction by an
external field. In an oblate condensate, vortex line can be
approximated as straight. This is not the case for a prolate
condensate. In that case, vortex lines are twisted. In the case of
short range contact interaction, an off-axis vortex in a BEC was
studied in detail \cite{jackson,lund,off1,off2,off3,off4}. We will
analyze the effects of the dipolar interaction on the physics of
an off-axis vortex. Specifically, we shall assume that a straight
singly quantized vortex line in an oblate trap is displaced from
the trap center of the $\ds{^{52}Cr}$ condensate with
transverse coordinates $\ds{x_0}$ and $\ds{y_0}$.\\
This paper is structured as follows. Sec. II reviews the TF
solution for a dipolar condensate. Section III investigates the
singly quantized straight vortex line in the presence of the
dipole-dipole interaction. The last section discusses the results.

\section{TF Solution}

In this section, we will review the TF solution for a dipolar gas.
Interparticle interaction potential in dipolar gases includes both
a short-range Van der Waals and a long-range dipole-dipole terms.
Because of the long-range character of the dipole-dipole
interaction, scattering properties at low energies are
significantly changed. In the case of a short-range interaction,
only the s-wave scattering is important at low energies. However,
in the case of a long-range interaction, all partial waves
contribute to scattering. Within the mean-field description of the
condensate, the interaction potential is well described by the
following model potential \cite{g1,g2,g3,g4,rev1}
\begin{equation}\label{de21sdgz}
V=g~\delta(\textbf{r})+\frac{d^2}{r^3}(1-3 \cos^2 \theta)~,
\end{equation}
where $\ds{g=\frac{4\pi \hbar^2a_s}{m}}$, $\ds{a_s}$ is the
scattering length, $d$ is the electric dipole moment (the results
are equally valid for magnetic dipoles), $\ds{\mathbf{r}}$ is the
vector connecting two dipolar particles and $\ds{\theta}$ is the
angle between $\ds{\mathbf{r}}$ and the dipole orientation. In
this study, we will suppose that the dipoles are polarized along
the z-axis.\\
For dipolar condensates, it is useful to introduce a dimensionless
parameter that measures the relative strength of the dipolar and
s-wave interactions
\begin{equation}\label{lrkhjt}
\varepsilon_{dd}\equiv\frac{C_{dd}}{3 g }
\end{equation}
where the coupling $C_{dd}=\mu_{0} \mu^{2}$. Chromium atoms posses
an anomalously large magnetic dipole moment $\mu_{Cr}=6 \mu_{B}$
($\mu_{B}$ is the Bohr magneton), while $\ds{^{87}Rb}$ has a
dipole moment equals to $\mu_{Rb}=1 \mu_{B}$ \cite{rev1}. It has
been shown in \cite{TF} that in the Thomas-Fermi limit a dipolar
BEC also is stable as long as $ 0< \varepsilon_{dd}<1 $.\\
Consider a dipolar BEC of $\ds{N}$ particles with mass $\ds{m}$
and electric dipole $\ds{d}$ oriented in the z-direction by a
sufficiently large external field. At sufficiently low
temperatures, the description of the ground state of the
condensate is provided by the solution of the Gross-Pitaevskii
(GP) equation
\begin{equation}\label{denklem1}
i\hbar \frac{\partial \Psi(\mathbf{r},t)}{\partial
t}=\left(-\frac{\hbar^2}{2m} \nabla^2 +V_{T}+g
|\Psi(\mathbf{r},t)|^2+d^2 \Phi_{dd}(\textbf{r})
\right)\Psi(\mathbf{r},t)~,
\end{equation}
where $\ds{V_{T}}$ is the trap potential
\begin{equation}\label{trappotential}
V_{T}=\frac{m}{2}\omega_{\perp}^{2} ~(\rho^{2}+\gamma^{2} z^{2})~,
\end{equation}
where $\ds{\rho^{2}=x^{2}+y^{2}}$ and $\ds{
\gamma\equiv\frac{\omega_{z}}{\omega_{\perp}}}$ is the trap aspect
ratio and $\ds{\Phi_{dd}(\textbf{r})=\int d^3\mathbf{r^{\prime}}
\frac{1-3 \cos^2 \theta}{|\mathbf{r}-\mathbf{r^{\prime}}|^3}
|\Psi(\mathbf{r^{\prime}},t)|^2}$ is the mean-field potential due
to dipole-dipole interactions.\\
The equation (\ref{denklem1}) is an integro-differential equation
since it has both integrals and derivatives of an unknown wave
function. This equation can be solved analytically if we assume
that the zero-point kinetic energy associated with the density
variation becomes negligible in comparison to both the trap energy
and the interaction energy between atoms. In this case, the
kinetic energy term can be omitted in the equation. This
approximation is known as TF approximation. Eberlein et al. showed
that a parabolic density profile remains an exact solution for an
harmonically trapped vortex-free dipolar condensate in the TF
limit \cite{TF}.
\begin{equation}\label{nbg}
n_{bg}(\textbf{r})=n_{0}
(1-\frac{\rho^{2}}{{R}^{2}}-\frac{z^{2}}{{L}^{2}})
\end{equation}
In the absence of dipolar interaction, the condensate aspect
ratio, $\ds{\kappa\equiv\frac{R}{L}}$, and the trap aspect ratio,
$\ds{\gamma}$, match. However, the presence of the dipolar
interaction changes the condensate aspect ratio. It satisfies the
following equation
\begin{equation}\label{gffjkký}
3\kappa^{2}\varepsilon_{dd}[(\frac{\gamma^{2}}{2}+1)\frac{f(\kappa)}{1-\kappa^{2}}-1]+(\varepsilon_{dd}-1)(\kappa^{2}-\gamma^{2})=0
\end{equation}
where $\ds{f(\kappa)}$ for oblate case, ($\kappa>1$), is given by
\begin{equation}\label{lnjbhf}
f(\kappa)\equiv\frac{2+\kappa^{2}(4-6~
\frac{\arctan{\sqrt{\kappa^{2}-1}}}{\sqrt{\kappa^{2}-1}}~)}{2
(1-\kappa^{2})}
\end{equation}
In the TF regime, the mean-field potential integral due to
dipole-dipole interactions, $\ds{\Phi_{dd}(\textbf{r})}$, can be
evaluated in the spheroidal coordinates \cite{TF}. The result can
be expressed in cylindrical coordinates.
\begin{equation}\label{fidd}
\Phi_{dd}^{bg}(\rho,z)=\frac{n_{0} C_{dd}}{3} \left(
\frac{\rho^{2}}{{R}^{2}}-\frac{2
z^{2}}{{L}^{2}}-f(\kappa)(1-\frac{3}{2} \frac{\rho^{2}-2
z^{2}}{{R}^{2}-{L}^{2}})   \right)
\end{equation}
Let us now write the energy expression for a dipolar gas. The
total energy for a dipolar gas can be written as
\begin{equation}\label{totale}
E_{tot}=E_{kin}+E_{trap}+E_{sw}+E_{dd}~,
\end{equation}
where $\ds{E_{kin}}$ is the kinetic energy
\begin{equation}\label{kinetice}
E_{kin}=-\frac{\hbar^{2}}{2 m} \int d^{3} r \Psi^{\ast}
(\textbf{r}) \nabla^{2}\Psi(\textbf{r})~,
\end{equation}
$\ds{E_{trap}}$ is the trap energy
\begin{equation}\label{trape}
E_{trap}=\int d^{3}r |\Psi(\textbf{r})|^{2}V_{T}~,
\end{equation}
$\ds{E_{sw}}$ is the energy due to the short range interaction
\begin{equation}\label{swe}
E_{sw}=\frac{g}{2}\int d^{3}r |\Psi(\textbf{r})|^{4}~,
\end{equation}
and $\ds{E_{dd}}$ is the energy due to the long range
dipole-dipole interaction
\begin{equation}\label{ddghfe}
E_{dd}=\frac{1}{2} \int d^{3}r d^{3}
r^{\prime}|\Psi(\textbf{r})|^{2}U_{dd}(\textbf{r}-\textbf{r}^{\prime})|\Psi(\textbf{r}^{\prime})|^{2}=\frac{1}{2}\int
d^{3}r~ n(\textbf{r}) \Phi_{dd}(\textbf{r})~.
\end{equation}
In the TF approximation, the kinetic energy term (\ref{kinetice})
is neglected. The total energy  associated with the vortex-free
Thomas-Fermi solution is given by \cite{TF}
\begin{equation}\label{þkjýyt}
E_{tot}=\frac{N}{14} m \omega_{x}^{2}
{R}^{2}(2+\frac{\gamma^{2}}{\kappa^{2}})+\frac{15}{28 \pi}
\frac{N^{2} g }{{R}^{2} {L}}\left(1-\varepsilon_{dd}
f(\kappa)\right)
\end{equation}

\section{Dipolar Condensate with a Vortex}

Consider a single straight vortex line at a position $\ds{\rho_0}$
along the z-axis. In this case, the wave function, normalized to
the total number of atoms $\ds{\int|\Psi|^2 d^3 r=N}$, is given by
\begin{equation}\label{dfh74yu}
\Psi(\rho, \phi, z)=\sqrt{n(\textbf{r})}~e^{iS(\rho,\rho_0)}~,
\end{equation}
where $\ds{n(\textbf{r})\equiv|\Psi(\textbf{r})|^{2}}$ is the
density. The expression of the phase $\ds{S(\rho,\rho_0)}$
characterizing the circulating flow around the vortex line is
given by \cite{theory}
\begin{equation}\label{sss}
S(\rho,\rho_0)=\arctan\left(\frac{y-y_0}{x-x_0}\right)
\end{equation}
where $\ds{\rho_0^2=x_0^2+y_0^2}$. The corresponding irrotational
flow velocity is given by
\begin{equation}\label{velocity}
\vec{v}=\frac{\hbar}{m}\nabla S(\rho,\rho_0)
\end{equation}
There is a singularity on the vortex line, $\ds{\rho=\rho_0}$,
where the velocity diverges. However, the particle current
density, $\ds{\vec{J}=n \vec{v}}$, vanishes as
$\ds{\rho\rightarrow\rho_0}$. When a quantized vortex is present
at the position $\ds{\rho_0}$, the density drops to zero at the
center of the vortex core whose size is determined by the
parameter $\ds{\beta}$. For a centered vortex in a BEC without
dipole-dipole interaction, the parameter $\ds{\beta}$ is given by
\begin{equation}\label{53hd}
\frac{\beta_{sw}}{R}=\left(\frac{d_{\perp}}{R}\right)^2~.
\end{equation}
where $\ds{d_{\perp}=\sqrt{\hbar/m\omega_{\perp}}}$ is the mean
oscillator strength. The TF limit holds when $\ds{R}$ is large
compared to $\ds{d_{\perp}}$. The TF length scale reads
$\ds{\beta<<d_{\perp}<<R}$. The vortex core size increases with
$\ds{\rho_0}$. The parameter $\ds{\beta_{sw}(\rho_0)}$
characterizing the small vortex core at the position $\ds{\rho_0}$
in a BEC without the dipole dipole interaction is \cite{lund}
\begin{equation}\label{53hd}
\beta_{sw}(\rho_0)=\frac{\beta_{sw}}{\sqrt{1-\rho_0^2/R^2}}~.
\end{equation}
Having written the expression of the phase, for a straight
off-center vortex, let us now find the density
$\ds{n(\textbf{r})}$. The repulsive interactions and the repulsive
dipolar interaction (for oblate case) significantly expand the
condensate, so that the kinetic energy associated with the density
variation becomes negligible compared to the trap energy and
interaction energies. In the TF regime, the density profile of a
condensate with a straight off-axis vortex line at $\ds{\rho_0 }$
is given by \cite{jackson}
\begin{equation}\label{density}
n(\textbf{r})=n_{0}\left(1-\frac{\rho^{2}}{R^{2}}-\frac{z^{2}}{L^{2}}\right)
\left(\frac{| \rho-\rho_0   |^2}{| \rho-\rho_0
|^2+\beta^{2}}\right)
\end{equation}
where $\ds{n(\textbf{r})=0}$ when the right hand side is negative
and $\ds{\beta}$, $\ds{R}$, and $\ds{L}$ are variational
parameters that describe the size of the vortex core, and the
radial and the axial sizes, respectively. Note that the density
function (\ref{density}) behaves like $\ds{| \rho-\rho_0
|^2/\beta^2}$ when $\ds{\rho<<\beta}$ and like
$\ds{(1-\frac{\rho^{2}}{{R}^{2}}-\frac{z^{2}}{{L}^{2}})}$ when
$\ds{\rho>>\beta}$. These parameters will be calculated by
minimizing the energy functional. The central density $n_{0}$ can
be found using the normalization condition
\begin{equation}\label{n0}
n_{0}=\frac{N \kappa}{32 \pi R^5} \left(  60{R}^2
+25{\bar{\beta}}^{2} \left (4{R}^2(3\ln({\frac{2}{\bar{\beta}}})-4
)+9 \rho_{0}^{2} (3-2\ln({\frac{2}{\bar{\beta}}})) \right) \right)
\end{equation}
We don't include an image vortex because the form of the TF
condensate density ensures that the particle current density
vanishes at the surface.\\
Let the total angular momentum for a singly quantized vortex line
along the trap axis at the position $\ds{\rho_0}$ be $\ds{L_z}$,
($\ds{L_z=m\int{rv_{\phi}n(\textbf{r})d^3\textbf{r}}}$). Then the
corresponding energy of the system in the rotating frame is
$\ds{E^{\prime}=E-\Omega L_{z}}$, where $\ds{E}$ is the energy in
the non-rotating frame. If we denote the energy of the BEC in its
ground state without a vortex by $\ds{E_0}$ and the extra energy
needed to generate a vortex by $\ds{\Delta E}$, then
$\ds{E=E_0+\Delta E}$. We can now write the energy of the vortex
state in the rotating frame as
\begin{equation}\label{ryhfdx}
E^{\prime}=E_{0}+\Delta E-\Omega L_{z}
\end{equation}
A vortex is generated if $\ds{E^{\prime}}$ is smaller than
$\ds{E_0}$. In other words, a vortex is formed above a certain
critical value of the rotation frequency. The critical rotational
velocity is given by
\begin{equation}\label{þkýjgtr}
\Omega_{c}=\frac{\Delta{E}}{L_{z}}
\end{equation}
It should be noted that a vortex lattice starts to appear when the
rotation frequency is further increased.\\
We proceed by minimizing the total energy with
respect to the three variational parameters, $\ds{R}$,
$\ds{\kappa}$ and $\ds{\beta}$. Let us now calculate the kinetic,
trap, s-wave and dipole-dipole interactions energies separately.
Since $\ds{\beta}$ is small, we will neglect terms of order
$\ds{\beta^3}$ and higher. The energy integral can be evaluated
analytically up to the second order of $\ds{\rho_0}$. Below we
will obtain analytical expression for small $\ds{\rho_0}$. In the
following section, we will perform numerical calculations for
large $\ds{\rho_0}$ in the TF limit.\\
Let us firstly obtain the kinetic energy
\begin{equation}\label{mkhuhg}
E_{kin}=\frac{ \hbar^{2} \pi n_{0} R}{9\kappa m}\left(-22
+12(1+3\bar{\beta}^2) \ln(\frac{2}{\bar{\beta}})-27 \bar{\beta}^2
+18\left(2-(1+2\bar{\beta}^2 )\ln(\frac{2}{\bar{\beta}})
\right){\bar{\rho}_{0}}^{2}\right)
\end{equation}
where we have defined
\begin{equation}\label{kmjklnmln}
\bar{\beta}=\frac{\beta}{R}~,~~\bar{\rho}_{0}=\frac{\rho_{0}}{R}~.
\end{equation}
Using the expression (\ref{trape}), the trapping energy is
straightforwardly evaluated to be
\begin{equation}\label{?lojj?}
E_{trap}=\frac{ 4 \pi  m  n_{0} \omega_{x}^{2} {R}^{5}}{15
\kappa^{3}}\left(\frac{1}{7}(2
\kappa^{2}+\gamma^{2})-\frac{{\bar{\beta}}^{2}}{15}(15\kappa^{2}-\gamma^{2}(23-15\ln({\frac{2}{\bar{\beta}}}))
)+\frac{5 {\bar{\beta}}^{2} }{12}\left(28 \kappa^{2}-11
\gamma^{2}+6\ln({\frac{2}{\bar{\beta}}})(\gamma^{2}-2\kappa^{2})
\right)\bar{\rho}_{0}^{2}\right)
\end{equation}
In the similar way, the formula (\ref{swe}) yields the s-wave
interaction energy
\begin{equation}\label{i??kjl}
E_{sw}= \frac{8  \pi g {R}^{3} n_{0}^{2}}{15
\kappa}\left(\frac{2}{7}+\frac{107}{15}
{\bar{\beta}}^{2}-4{\bar{\beta}}^{2}\ln({\frac{2}{\bar{\beta}}})
-\frac{5\bar{\beta}^{2}}{6}\left(25-12\ln({\frac{2}{\bar{\beta}}})\right){\bar{\rho}_{0}}^{2}\right)
\end{equation}
Let us now calculate the dipole-dipole interaction energy. Since
$\ds{\beta}$ is small, the dipolar energy function can be
approximated as \cite{TFvortex}
\begin{equation}\label{dde2}
E_{dd}\approx\frac{1}{2}\int d^{3} r n _{bg}(\textbf{r})
\Phi_{dd}^{bg}(\textbf{r})+\int d^{3} r n
_{v}(\textbf{r})\Phi_{dd}^{bg}(\textbf{r})
\end{equation}
where $\ds{n_{bg}}$ was defined in (\ref{nbg}) and $\ds{n_{v}}$ is
defined as
\begin{equation}\label{nv}
n_{v}(\textbf{r})=-n_{0}~ \frac{\beta^{2}}{| \rho-\rho_0
|^2+\beta^{2}}\left(1-\frac{\rho^{2}}{{R}^{2}}-\frac{z^{2}}{{L}^{2}}\right)~.
\end{equation}
Note that
$\ds{n(\textbf{r})=n_{bg}(\textbf{r})+n_{v}(\textbf{r})}$. Hence,
the dipolar interaction energy becomes $\ds{E_{dd}=E_{0dd}+\Delta
E_{dd}}$, where
\begin{equation}\label{ll56jkl}
E_{0dd}=  \frac{4 \pi g \varepsilon_{dd} R^{3} n_{0}^{2}}{225
\kappa}
\left(-\frac{60}{7}f(\kappa)+{\bar{\beta}}^2\left(60\ln({\frac{2}{\bar{\beta}}})-122+
\frac{f(\kappa)}{\kappa^{2}-1} (62-245 \kappa^{2}+30 (5
\kappa^{2}-2)\ln({\frac{2}{\bar{\beta}}}))\right)\right)
\end{equation}
\begin{equation}\label{78564}
\Delta E_{dd}=\frac{2 \pi g \varepsilon_{dd} R^3 {\bar{\beta}}^2
n_{0}^{2}}{9 \kappa} \left(50-24
\ln({\frac{2}{\bar{\beta}}})+\frac{f(\kappa)}{\kappa^{2}-1}(6+69
\kappa^{2}-36 \kappa^{2}
\ln({\frac{2}{\bar{\beta}}}))\right){\bar{\rho}_{0}}^{2}
\end{equation}
We have obtained the energy expressions up to the second order of
fractional vortex displacement, $\ds{{\bar{\rho}_{0}}}$. Note that
the central density $\ds{n_0}$ in these expressions also includes
$\ds{{\bar{\rho}_{0}}^{2}}$ (\ref{n0}). Up to the order of
$\ds{\bar{\beta}^2}$, they
agree with the results \cite{TFvortex} in the limit $\ds{\rho_0{\rightarrow}~0}$.\\
As can be seen, the kinetic energy decreases with $\ds{\rho_0}$.
The kinetic energy goes to zero as $\ds{\rho_0{\rightarrow}R}$
since TF density vanishes at the surface. Note that the
description of a vortex close to the boundary is outside the scope
of the present approach, since TF approach doesn't work close to
the surface. The dipole-dipole interaction increases with
$\ds{\rho_0}$ for an oblate trap while decreases with
$\ds{\rho_0}$ for a prolate trap.The kinetic energy depends on
$\ds{{\bar{\rho}_0}^2}$ while the dipolar, trap and the s-wave
interaction energies depend on
$\ds{{\bar{\beta}}^2~{\bar{\rho}_0}^2}$.\\
Before embarking on a specific example, let us study the energy
expressions qualitatively for an oblate trap. Firstly, let us
investigate roughly how the total energy is distributed among
kinetic, dipolar, trap and s-wave interaction energies. The ratio
between the kinetic energy and the trap energy is of order
$\ds{{\bar{\beta}}^2}$;
$\ds{E_{kin}{\approx}~{\bar{\beta}}^2E_{trap}}$. The trap and the
s-wave interaction energies are comparable to each other;
$\ds{E_{sw}{\approx}~E_{trap}}$. The ratio between dipolar and the
s-wave interaction energies is of order $\ds{\epsilon_{dd}}$; $
\ds{E_{dd}{\approx}~{\epsilon_{dd}}E_{sw}}$.\\
Secondly, let us investigate how the excess energy
$\ds{{\Delta}E}$ needed to generate a vortex is distributed.
Consider first a central vortex, $\ds{\rho_0=0}$. The excess
energy for the trap, dipolar and s-wave interaction energies vary
as $\ds{{\bar{\beta}}^2}$. However the excess kinetic energy is of
order the kinetic energy,
$\ds{{\Delta}E_{kin}{\approx}E_{kin}{\approx}~{\bar{\beta}}^2E_{trap}}$.
Hence, $\ds{{\Delta}E_{kin}{\approx}~{\Delta}E_{trap}}$. We
emphasize that the excess energy for dipole-dipole and s-wave
interaction energies are negative. Hence, the effect of increasing
dipole moment and scattering length is to decrease the critical
angular velocity $\ds{\Omega_c}$. The relations between the excess
energies for dipolar, trap and s-wave terms are given by $
\ds{{\Delta}E_{dd}{\approx}~{\epsilon_{dd}}~{\Delta}E_{sw}}$ and
$\ds{{\Delta}E_{trap}{\approx}~-{\Delta}E_{sw}}$.\\
Finally, let us mention how the energy changes with the position
of an off-axis vortex, $\ds{\rho_0}$. The kinetic energy and the
trap energy decrease with $\ds{\rho_0}$ while the dipole-dipole
and the s-wave interaction energies increase. Furthermore, the
total energy decreases with $\ds{\rho_0}$.\\
Let us briefly study qualitatively the energy expressions for a
prolate trap with a straight vortex line. If we neglect vortex
bending effect, we can use the above energy expressions. In this
case, the dipolar interaction energy is negative. However, the
excess energy for the dipolar interaction is positive. Hence, the
effect of increasing dipole moment for a prolate trap is to
increase the critical angular velocity $\ds{\Omega_c}$. Finally,
the dipole-dipole interaction energy decreases with increasing
$\ds{\rho_0}$. As a result, the effect of dipolar interaction is
to repel an off axis vortex away from the trap center for a
prolate trap while attract it to the trap center for an oblate trap.\\
In what follows, we will give an explicit example for a straight
off-axis vortex for an oblate trap.

\section{Results}

\begin{figure}[htp]
\includegraphics[width=6cm]{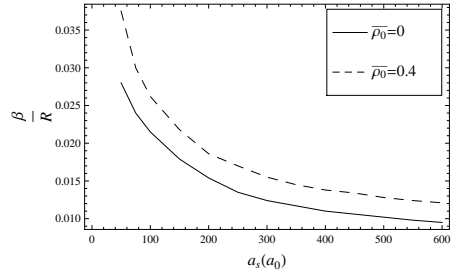}
\caption{\label{fig:7}The fractional vortex core size for a cental
($\ds{\bar{\rho}=0}$) and an off-axis ($\ds{\bar{\rho}=0.4}$)
vortices versus the scattering length for a dipolar BEC with a
vortex in an oblate trap with $\ds{\gamma=5}$. The scattering
length is measured in units of Bohr radius, $\ds{a_0}$. }
\end{figure}

\begin{figure}[htp]
\includegraphics[width=6cm]{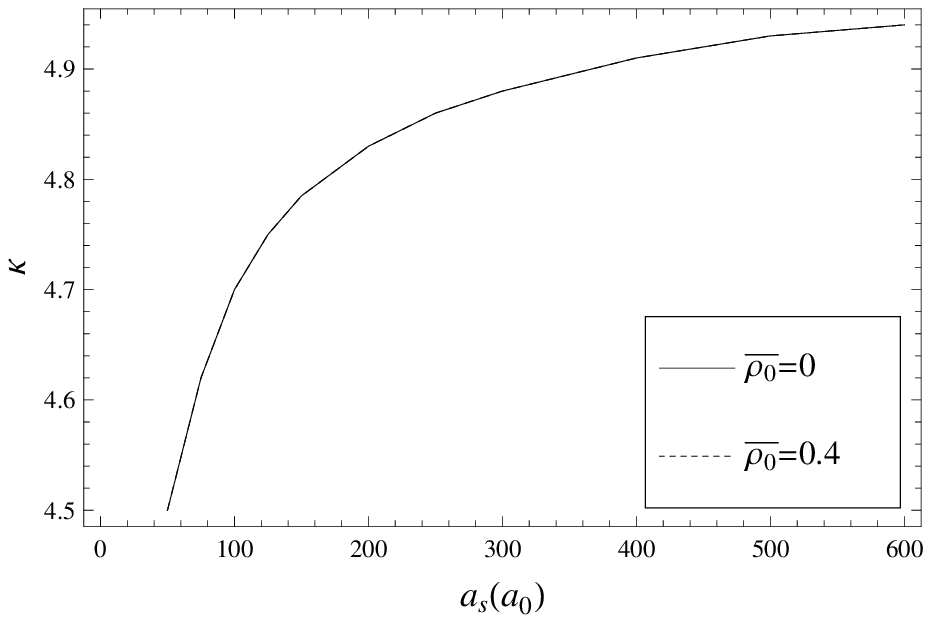}
\includegraphics[width=6cm]{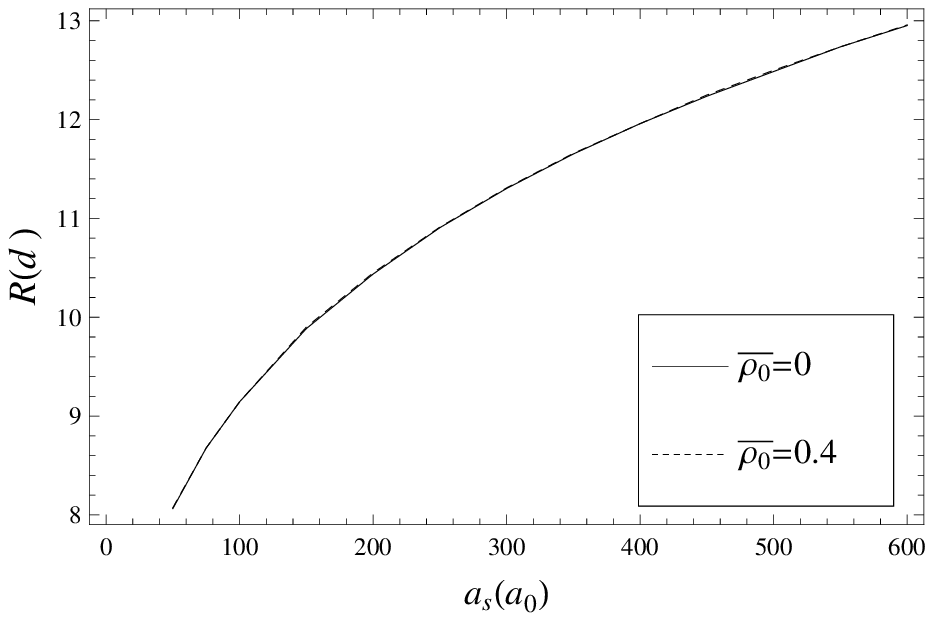}
\caption{\label{fig:6}For $\ds{\bar{\rho}=0}$ and
$\ds{\bar{\rho}=0.4}$, the aspect ratio (left) and the radial size
(right) of a dipolar BEC with a vortex in an oblate trap with
$\ds{\gamma=5}$ as a function of the scattering length. The
scattering length is measured in units of Bohr radius,
$\ds{a_0}$.}
\end{figure}

\begin{figure}[htp]
\includegraphics[width=6cm]{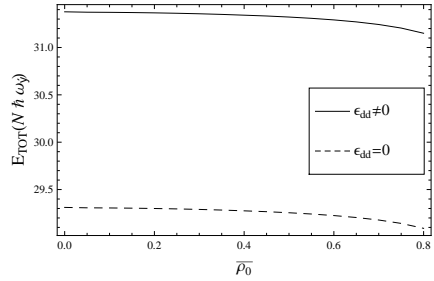}
\caption{\label{fig:1}The total energy of a non-rotating dipolar
BEC in an oblate trap with $\gamma=5$ for the fixed scattering
length $\ds{a_{s}=100a_{0}}$ as a function of vortex displacement.
The solid curve indicates both dipolar and s-wave interaction with
$\epsilon_{dd}=0.15$ while the dashed curve indicates only s-wave
interaction. }
\end{figure}

\begin{figure}[htp]
\includegraphics[width=6cm]{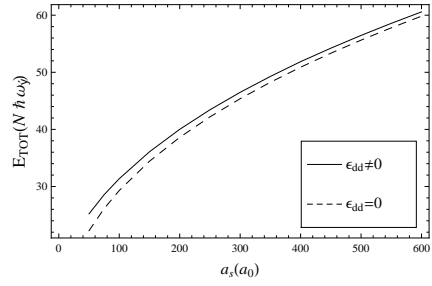}
\caption{ \label{fig:3}The total energy of a non-rotating dipolar
BEC with a vortex in an oblate trap ($\ds{\gamma=5}$) with
$\ds{\rho_{0}=0.2R}$ as a function of the scattering length (in
units of Bohr radius). The solid (dashed) curve is for the
condensate with both dipolar and s-wave interaction (only s-wave
interaction). }
\end{figure}

\begin{figure}[htp]
\includegraphics[width=6cm]{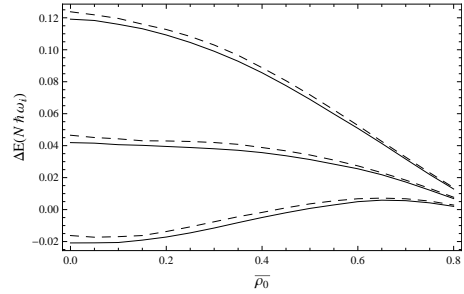}
\caption{\label{fig:silinecek}The increased energy
$\ds{{\Delta}E}$ in units of $\ds{N\hbar\omega_{\perp}}$ in the
rotating frame associated with the presence of an off-axis
straight vortex as a function of a fractional vortex displacement
in an oblate trap. The solid (dashed) curves correspond to
$\ds{\epsilon_{dd}=0.15}$ $\ds{(\epsilon_{dd}=0)}$. Different
curves represent different fixed values of the external angular
velocity $\ds{\Omega}$. The top solid and dashed curves
corresponds to $\ds{\Omega=0}$, where $\ds{\Omega}$ increases as
one moves towards the lowest curve with $\ds{\Omega=0.08~
\omega_{\perp}}$ $\ds{\Omega=0.14 ~\omega_{\perp}}$.}
\end{figure}

\begin{figure}[htp]
\includegraphics[width=6cm]{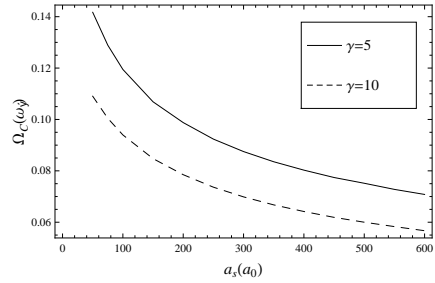}
\caption{\label{fig:8} The critical angular velocity of a
condensate  with a vortex for $\ds{\gamma=5}$ and $\ds{\gamma=10}$
as a function of the scattering length. $\ds{\Omega_{C}}$ is
measured in units of $\ds{\omega_{\perp}}$}
\end{figure}

We will study a dipolar BEC with a single vortex in an oblate trap
including 150000  $\ds{^{52}Cr}$ atoms. We take the numerical
values used in the reference \cite{TFvortex} to compare the
off-center vortex to the central vortex. The trap frequencies are
$\omega_{\perp}=2 \pi\times 200 $~ rad/s and $\omega_{z}=2 \pi
\times1000~ rad/s$ for $\gamma=5$. The harmonic oscillator length
of the trap along the radial direction is $\ds{d_{\perp}=0.986 \mu
m}$. The magnitude of the magnetic dipole interaction for
$\ds{^{52}Cr}$ is $C_{dd}=\mu_{0} (6 \mu_{B})^{2}$. For small
values of $\ds{\rho_0}$, we will use the analytical results
obtained in the previous section. For large values $\ds{\rho_0}$,
numerical computation within the TF limit will be performed.\\
Let us firstly analyze three variational parameters, $\ds{\beta}$,
$\ds{\kappa}$ and $\ds{R}$. The density of the condensate drops to
zero at the center of the vortex core whose size is equal to
$\ds{\beta}$. It is very small compared to the radial size of the
condensate. The smallness of $\ds{\bar{\beta}}$ ensures that the
vortex affects the density only the immediate vicinity of the
core. Fig-\ref{fig:7} depicts the fractional vortex core size,
$\ds{\bar{\beta}=\beta/R}$, versus the scattering length. The
solid curve corresponds to a central vortex, while the dashed
curve to an off-center vortex with $\ds{\rho_{0}=0.4 R}$. The
parameter $\ds{\bar{\beta}}$ is bigger in the presence of an
off-axis vortex. As can be seen from the figure, the fractional
vortex core size decreases with increasing scattering length. This
can be understood as follows. The radial size increases as
scattering length is enlarged. The vortex core size is inversely
proportional to radial size. So, we conclude
that $\ds{\bar{\beta}}$ decreases with $\ds{a_s}$. \\
Similarly, Fig-\ref{fig:6} show the aspect ratio $\kappa$ (left)
and the radial size (right) of the condensate $\ds{R}$ versus the
scattering length, respectively for $\rho_{0}=0$ and
$\ds{\rho_{0}=0.4 R}$. Contrary to the case of vortex core size
$\ds{\beta}$, the parameters $\ds{\kappa}$ and $\ds{R}$ don't
change appreciably with $\ds{\rho_{0}}$ when $\ds{a_s>50 a_0}$,
where $\ds{a_0}$ is Bohr radius. Hence, the curves lie on top of each other in Fig-\ref{fig:6}.\\
Having discussed the three variational parameters, let us now
study the total energy of a dipolar condensate. Figures
\ref{fig:1} and \ref{fig:3} show the total energy as a function of
vortex position for fixed $\ds{a_s=100 a_{0}}$ and scattering
length for fixed $\ds{\bar{\rho_0}=0.2}$ in a non-rotating oblate
trap ($\ds{\Omega=0}$), respectively. The solid curve corresponds
to a condensate with s-wave plus dipolar interactions while the
dashed line corresponds to a condensate with pure s-wave
interaction. The total energy is bigger when
$\ds{\epsilon_{dd}{\neq}0}$. This is because the dipole-dipole
interaction energy is positive in an oblate trap. The total energy
of the system attains a maximum when $\ds{\rho_0=0}$ for both
cases. As the off-axis vortex moves to the edge of the condensate,
the total energy decreases. More specifically, the kinetic and
trap energies decrease with $\ds{\bar{\rho_0}}$ while the dipolar
and s-wave interaction energies increase with $\ds{\bar{\rho_0}}$.
In fact, higher than a specific value of $\ds{\epsilon_{dd}}$,
dipolar interaction becomes more dominant, so total energy
increases with $\ds{\bar{\rho_0}}$. We calculate that this happens
when $\ds{\epsilon_{dd}>1}$. As mentioned in \cite{TFvortex},
however, the condensate enters an instability region when
$\ds{\epsilon_{dd}>1}$. As can be seen from the figure
\ref{fig:3}, the energy differences between the two cases
decreases when the scattering length is increased. This is because
$\ds{\epsilon_{dd}}$ is decreased with increasing scattering
length (\ref{lrkhjt}).\\
For the investigation of the vortex generation, not the total
energy but the excess energy $\ds{{\Delta}E}$ associated with the
presence of an off-axis straight vortex is more important.
Fig-\ref{fig:silinecek} compares the excess energy of the
condensates with $\ds{\epsilon_{dd}=0.15}$ (solid curves) and
$\ds{\epsilon_{dd}=0}$ (dashed curves) as a function of a
fractional vortex displacement. Different curves represent
different fixed values of the external angular velocity
$\ds{\Omega}$. The top solid and dashed curves correspond to
$\ds{\Omega=0}$, where $\ds{\Omega}$ increases as one moves
towards the lowest curve with $\ds{\Omega=0.08~ \omega_{\perp}}$
and $\ds{\Omega=0.140 ~\omega_{\perp}}$. Note that the critical
rotation frequency is $\ds{\Omega_c=0.124~ \omega_{\perp}}$
$(\ds{\Omega_c=0.119~ \omega_{\perp})}$ when
$\ds{\epsilon_{dd}=0}$ $(\ds{\epsilon_{dd}=0.15})$. As can be seen
from the figure, the dipolar interaction lowers $\ds{{\Delta}E}$
compared to the pure contact interaction. It is of great
importance to note that although dipolar interaction is positive
for an oblate trap, the excess dipolar energy is negative. As
$\ds{\bar{\rho}_{0}}$ is increased, the curves for
$\ds{\epsilon_{dd}=0.15}$ and $\ds{\epsilon_{dd}=0}$ start to
coincide.  The top two curves show that the excess energy
$\ds{{\Delta}E}$ decreases monotonically with increasing
$\ds{\bar{\rho}_{0}}$, with negative curvature at
$\ds{\bar{\rho}_{0}=0}$. So, a central vortex is unstable to
infinitesimal displacements. The presence of dissipation will move
an off-axis vortex toward the edge of the condensate. If the trap
is rotated with angular velocity, $\ds{\Omega}$, then the energy
of a vortex decreases. Inspection of Fig. \ref{fig:silinecek}
reveals that with increasing rotation speed, the function
$\ds{{\Delta}E}$ flattens. At a special value of rotation
frequency, $\ds{\Omega_m}$, curvature of the function
$\ds{{\Delta}E}$ becomes zero at $\ds{\bar{\rho}_{0}=0}$. Hence,
above an angular velocity $\ds{\Omega_m}$, the vortex attains a
local minimum. The central position is not globally stable but
locally stable. One of the results of this paper is that the
presence of dipolar interaction lowers $\ds{\Omega_m}$ for an
oblate trap. Let us look at the lowest curves in Fig.
\ref{fig:silinecek}. In this case, appearance of a vortex becomes
energetically favorable since $\ds{{\Delta}E<0}$. The central
vortex is both locally and globally stable relative to the
vortex-free state. A vortex initially placed off-center will
follow a path of constant energy under the action of the Magnus
force, which is proportional to the gradient of the energy in the
radial direction. The precession velocity of a displaced vortex of
a nonrotating trap increases with the vortex displacement. Hence,
a vortex near the surface precesses more rapidly than one near the
center. Another result of this paper is that the precession
velocity of a displaced vortex is lowered in the presence of the
dipolar interaction in an oblate trap. On the contrary, it is
raised in a prolate trap (ignoring vortex bending effect). Note
that the precession velocity around the center for a nonrotating
trap, $\ds{\omega}$, can be calculated using
$\ds{\omega=\frac{{\partial}E}{{\partial}L}=\frac{{\partial}E/{\partial}{\bar{\rho_0}}}{{\partial}L/{\partial}{\bar{\rho_0}}}}$,
where $E$ is the energy and $L$ is the angular momentum
\cite{theory,lund}. For a condensate in rotational equilibrium at
angular velocity $\ds{\Omega}$, the original precession frequency
is altered to $\ds{\omega\rightarrow{\omega}-\Omega}$ \cite{theory}.\\
Finally, in Fig-\ref{fig:8}, we have examined the critical angular
velocity of the condensate for $\gamma=5$ and $\gamma=10$. The
critical angular velocity above which a vortex state is
energetically favorable depends on $\gamma$. As can be seen,
$\Omega_{c}$ increases with decreasing $\gamma$. For stirring
frequencies below $\ds{\Omega_c}$, no vortex can be nucleated. The
presence of dipole dipole interaction decreases $\Omega_{c}$ for an oblate trap.\\
We have found that  $\Omega_{c}$, $\Omega_{m}$ and precession
velocity decrease (increase) in an oblate (a prolate) trap. This
can be understood simply as follows. The dipolar mean field
potential has a parabolic profile $\ds{\Phi^{bg}_{dd}(\rho,z)=
n_0C_{dd}/3L^2(1.21-0.04 \rho^2-1.83 z^2)}$ when
$\ds{\epsilon_{dd}=0.15}$ and $\ds{\gamma=5}$ (\ref{fidd}). This
potential has the same inverted parabola shape as in the case of
contact interactions. So, we conclude that there is a similarity
between dipolar and non-dipolar BEC in the TF regime. The
difference is in the expressions for the radial and axial size. It
is well known that the contact interaction with positive
scattering length decreases the critical angular frequency
$\Omega_{c}$ ($\Omega_{c}=\omega_{\perp}$ for a noninteracting
trapped gas). In the similar way, $\Omega_{m}/\omega_{\perp}$ and
precession velocity decrease with increasing scattering length.
So, we conclude that inclusion of dipolar interaction in an oblate
trap reduces $\Omega_{c}$, $\Omega_{m}$ and precession velocity in
the TF regime. Furthermore, if we ignore vortex bending effect,
the mean field dipolar potential for a prolate trap has the same
form as the mean field contact potential with negative scattering
length. In contrast to the case for repulsive interactions,
$\Omega_{c}$, $\Omega_{m}$ and precession velocity  increase in
the presence of attractive contact interactions. Analogously, we
conclude that dipolar interactions increase them in a prolate trap.\\
In this paper, an off-axis vortex lines in an oblate
$\ds{^{52}Cr}$ BEC polarized along the $\ds{z}$ direction have
been studied. The effects of the dipolar interaction on the
physics of an off-axis vortex have been analyzed. It was shown
that the condensate aspect ratio, $\ds{\kappa}$, and the radial
size, $\ds{R}$, remain almost the same in the presence of an
off-axis vortex when $\ds{a_s>50 a_0}$. The dipolar interaction
raises (lowers) the total energy in an oblate (a prolate) trap. On
the contrary, the excess dipolar energy needed to generate a
vortex decreases (increases) in an oblate (a prolate) trap. It was
found that the angular velocity $\ds{\Omega_m}$ representing the
onset of metastability and the critical angular velocity of the
condensate $\ds{\Omega_c}$ are lowered (raised) in an oblate (a
prolate) trap. Finally, it was proven that the effect of the
dipole-dipole interaction is to lower (raise) the precession
velocity of an off-axis straight vortex line around the center in
an oblate (a prolate) trap.

\newpage

\end{document}